\documentstyle[11pt]{article}
\begin{document} 

\noindent  {\bf A NEW CONDITION TO IDENTIFY 
ISOTROPIC \\ DIELECTRIC--MAGNETIC MATERIALS
DISPLAYING \\ NEGATIVE PHASE VELOCITY} 
\vskip 0.4cm
 
\noindent{ \bf Ricardo A. Depine$^1$ and Akhlesh Lakhtakia$^2$}\\
\vskip 0.2cm
{\sf
\noindent{$^1$ Grupo de Electromagnetismo Aplicado, Departamento de F\'{\i}sica }\\
\noindent{Facultad de Ciencias Exactas y Naturales, Universidad de Buenos Aires }\\
\noindent{Ciudad Universitaria, Pabell\'{o}n I}\\
\noindent{ 1428 Buenos Aires, Argentina}\\
\noindent{\em email: rdep@df.uba.ar}\\

\noindent{$^2$ Computational \& Theoretical Materials Sciences Group }\\
\noindent{Department of Engineering Science \& Mechanics }\\
\noindent{Pennsylvania State University }\\
\noindent{University Park, PA 16802-6812, USA}\\
\noindent{\em email: AXL4@psu.edu}
}

\noindent{\bf ABSTRACT:}
The derivation of a new condition for characterizing isotropic
dielectric--magnetic materials exhibiting negative phase velocity,
and the equivalence of that condition with previously derived conditions,
are presented. 


\vskip 0.4 cm

\noindent{\bf 1. INTRODUCTION}

Non--dissipative mediums with both simultaneously negative  permittivity 
and  permeability were first investigated by Veselago \cite{veselago} in 1968. 
These mediums support electromagnetic wave propagation in which the phase velocity is antiparallel 
to the direction of energy flow, and other unusual electromagnetic effects such as the reversal of  the Doppler effect and Cerenkov radiation. After the publication of Veselago's work, more than three decades went by
for the actual realization of artificial materials that are effectively isotropic, homogeneous, and possess negative real permittivity and permeability in some frequency range \cite{smith,LMW02}. 

When dissipation is included in the analysis, a general condition for the constitutive parameters of 
an isotropic dielectric--magnetic medium to have phase velocity directed oppositely to the power flow, was reported about two years 
ago  \cite{AEW}: Most importantly, according to
that condition, the real parts of both the permittivity and the
permeability need not be both negative.

In this communication, we derive a new condition for characterizing isotropic materials with negative phase velocity. Although this new condition looks very different from its 
predecessor \cite{AEW}, we also show here the equivalence between both conditions. 

\noindent{\bf 2. THE NEW CONDITION}

Let us consider a linear isotropic dielectric--magnetic medium characterized by 
complex--valued relative permittivity and relative permeability scalars 
$\epsilon=\epsilon_r + i  \epsilon_i$ and $\mu=\mu_r + i\mu_i$. An
$\exp(-i\omega t)$ time--dependence is implicit, with $\omega$ as
the angular frequency.

The wave equation 
gives the square of the complex--valued refractive index $n=n_r+in_i$ as
\begin{eqnarray}
n^2= \epsilon \mu \Rightarrow n_r^2 - n_i^2 + 2 i n_r n_i=
\mu_r \epsilon_r - \mu_i \epsilon_i
+i (\mu_i \epsilon_r + \mu_r \epsilon_i) \,. \;\;\;    \label{refind1}
\end{eqnarray}
The sign of $n_r$ gives the phase velocity direction, whereas the sign
of the real part of $n/\mu$, i.e.,
\begin{eqnarray}
{\rm Re} \Big(\frac{n}{\mu}\Big)=n_r \mu_r + n_i \mu_i \,,    \label{poynt1}
\end{eqnarray}
gives the directon of power flow \cite{AEW}.
Therefore, for this medium to have negative phase velocity and positive 
power flow, the following conditions should hold simultaneously:
\begin{eqnarray}
n_r < 0 \,, \;\;\;    \label{cond1} \\
n_r \mu_r + n_i \mu_i>0 \,, \;\;\;    \label{cond2}
\end{eqnarray}

Equation (\ref{refind1})  yields the biquadratic equation 
\begin{eqnarray}
n_r^4 - (\mu_r \epsilon_r - \mu_i \epsilon_i)\,n_r^2 - 
\frac{1}{4} (\mu_i \epsilon_r + \mu_r \epsilon_i)=0 \,. \;\;\;  \label{biq1}
\end{eqnarray}
This equation has only two real--valued solutions for  $n_r$, {\em viz.}, 
\begin{equation}
n_r=\pm \Big(\frac{|\epsilon| |\mu| + \mu_r \epsilon_r - \mu_i \epsilon_i}{2}\Big)^{1/2} 
\,. \;\;\;  \label{nrsqrd}
\end{equation}
Noting that the relation 
\begin{eqnarray}
\mu_i \epsilon_i - \mu_r \epsilon_r \, < \, 
\sqrt{(\mu_i \epsilon_i - \mu_r \epsilon_r)^2+
(\mu_i \epsilon_r + \mu_r \epsilon_i)^2}   \label{relat1}
\end{eqnarray}
holds for all values of the constitutive parameters $\epsilon_{r,i}$ and
$\mu_{r,i}$, we see that 
\begin{eqnarray}
0 \, < \, |\epsilon| |\mu| + \mu_r \epsilon_r - \mu_i \epsilon_i 
\,;  \label{relat2}
\end{eqnarray}
hence, the right  side of  (\ref{nrsqrd}) is always positive. 

As the negative square root  must be chosen
in  (\ref{nrsqrd}) in order to satisfy the
condition (\ref{cond1}), therefore
\begin{eqnarray}
&&n_r=-\frac{1}{\sqrt{2}}\,
\Bigl (|\epsilon|\,|\mu| + \mu_r \epsilon_r - 
\mu_i \epsilon_i\Bigr )^{1/2} \,,  \label{nrneg1} \\
&&n_i=-\frac{1}{\sqrt{2}}\,
\frac{\mu_i \epsilon_r + \mu_r \epsilon_i}
{\Bigl (|\epsilon|\,|\mu| + \mu_r \epsilon_r - \mu_i \epsilon_i\Bigr )^{1/2}}
\,.  \label{nineg1}
\end{eqnarray}
On using these expressions and   (\ref{poynt1}) in the
condition (\ref{cond2}), a condition for power flow and 
phase velocity in opposite directons is finally derived  as follows:
\begin{eqnarray}
\mu_r \Bigl (|\epsilon|\,|\mu| + \mu_r \epsilon_r - \mu_i \epsilon_i\Bigr )^{1/2} \,
+ \mu_i \,\frac{\mu_i \epsilon_r + \mu_r \epsilon_i}
{\Bigl (|\epsilon|\,|\mu| + \mu_r \epsilon_r - \mu_i \epsilon_i\Bigr )^{1/2}} 
\, < \, 0 \,.  \label{conditrad1}
\end{eqnarray}
This condition can be rewritten in the very simple form
\begin{eqnarray}
\epsilon_r |\mu| + \mu_r |\epsilon| <0  \,,  \label{conditrad2}
\end{eqnarray}
which is the chief contribution of this communication.

\noindent{\bf 3. EQUIVALENCE WITH  PREVIOUSLY DERIVED CONDITION}

The general condition  derived for the phase velocity to be 
oppositely directed to the power flow about two years ago \cite{AEW}
is  as follows:
\begin{eqnarray}
\Bigl (|\epsilon| - \epsilon_r \Bigr )\, \Bigl (|\mu| - \mu_r \Bigr ) 
\, > \, \epsilon_i \, \mu_i  \,.  \label{condital1}
\end{eqnarray}
Although it looks very different,   this condition, which can be rewritten as 
\begin{eqnarray}
\epsilon_r |\mu| + \mu_r |\epsilon| <  
|\epsilon| |\mu| + \mu_r \epsilon_r - \mu_i \epsilon_i  \,,  \label{condital2}
\end{eqnarray}
is completely equivalent to the new condition (\ref{conditrad2}). 

Clearly, if (\ref{conditrad2}) is satisfied, then, taking into account the 
validity of   (\ref{relat2}),  (\ref{condital2}) also is. To show that the reverse 
is also true, we start from (\ref{condital2}) and assume that (\ref{conditrad2}) does not hold. 
As the left  side of  the inequality (\ref{condital2}) is non--negative, 
squaring both sides does not change the sense of the inequality and we get
\begin{eqnarray}
\Bigl ( \epsilon_r |\mu| + \mu_r |\epsilon|\Bigr )^2 <  
\Bigl ( |\epsilon| |\mu| + \mu_r \epsilon_r - \mu_i \epsilon_i\Bigr )^2  \,.  \label{paso1}
\end{eqnarray}
Simplification of this inequality leads to
\begin{eqnarray}
\epsilon_i \mu_i \Bigl (
|\epsilon| |\mu| + \mu_r \epsilon_r - \mu_i \epsilon_i \Bigr ) \,<\,0
\,.  \label{paso2}
\end{eqnarray}
But causality dictates that $\epsilon_i \ge 0$ and $\mu_i \ge 0$; hence,
we must conclude that 
\begin{eqnarray}
|\epsilon| |\mu| + \mu_r \epsilon_r - \mu_i \epsilon_i  \,<\,0 \,,  \label{paso3}
\end{eqnarray}
in contradiction with Eq. (\ref{relat2}). Therefore, we must accept the validity of 
the condition (\ref{conditrad2}). This completes the demonstration of the equivalence between 
conditions (\ref{conditrad2}) and (\ref{condital1}).

\noindent{\bf 4. CONCLUDING REMARKS}

We note in passing that both conditions  (\ref{conditrad2}) and (\ref{condital1}) are 
also equivalent to the condition 
\begin{equation}
\label{Rupp}
\epsilon_r\mu_i + \mu_r\epsilon_i < 0\,,
\end{equation}
reported very recently \cite{LMW03}. This condition is due to R. Ruppin.

To conclude, we have here derived a simple new condition for the constitutive parameters 
of a linear isotropic dielectric--magnetic medium to have phase velocity opposite to the directon 
of power flow, and we have demonstrated its equivalence with previously derived conditions.

\vskip 0.4 cm

\noindent {\bf Acknowledgment.}
RAD acknowledges the support from Consejo Nacional de
Investigaciones Cient\'{\i}ficas y T\'ecnicas (CONICET) and Agencia Nacional
de Promoci\'on Cient\'{\i}fica y Tecnol\'ogica (ANPCYT-BID
802/OC-AR03-04457).  AL thanks the Mercedes Foundation for support.


\vskip 0.4cm

\begin{thebibliography}{99}
\bibitem{veselago} V. G. Veselago, 
The electrodynamics of substances with simultaneously negative values 
of $\epsilon$ and $\mu$,
Sov Phys Usp 10  (1968) 509--514. 

\bibitem{smith} 
R.A. Shelby, D. R. Smith  and S. Schultz, 
Experimental verification of a negative index of refraction, 
Science 292 (2001) 77--79.


\bibitem{LMW02} 
A. Lakhtakia. M.W. McCall and W.S. Weiglhofer,
Brief overview of recent developments on 
negative phase--velocity mediums, 
Arch Elektron \"Ubertrag 56 (2002) 407--410.

\bibitem{AEW} 
M.W. McCall, A. Lakhtakia and W. S Weiglhofer, 
The negative index of refraction demystified,
Eur J Phys 23  (2002) 353--359.


\bibitem{LMW03} 
A. Lakhtakia. M.W. McCall and W.S. Weiglhofer,
Negative phase--velocity mediums, In:
W.S. Weiglhofer and A. Lakhtakia (eds),
Introduction to complex mediums for optics and
electromagnetics,
SPIE  Press, Bellingham, WA, USA, 2003.



\end{thebibliography}
\enddocument